\documentclass[prd,aps,showpacs,preprintnumbers,amssymb]{revtex4}
\usepackage{graphicx}% Include figure files

\def\e3p{$\eta \rightarrow 3 \pi$}

\begin{document}

\title{%
\hfill{\normalsize\vbox{%
\hbox{\rm SU-4252-874}
 }}\\
{Simple two Higgs doublet model}}

\author{Renata Jora 
$^{\it \bf a}$~\footnote[1]{Email:
 cjora@phy.syr.edu}}

\author{Sherif Moussa
$^{\it \bf b}$~\footnote[2]{Email:
   sherif.moussa@uaeu.ac.ae          }}

\author{Salah Nasri
$^{\it \bf c}$~\footnote[3]{Email:
 nasri.salah@gmail.com}}
 
\author{Joseph Schechter
 $^{\it \bf d}$~\footnote[4]{Email:
 schechte@phy.syr.edu}}

\author{M. Naeem Shahid
$^{\it \bf e}$~\footnote[5]{Email:
   mnshahid@phy.syr.edu            }}

\affiliation{$^ {\bf \it a,d,e}$ Department of Physics,
 Syracuse University, Syracuse, NY 13244-1130, USA,}
\affiliation{$^ {\bf \it b}$ Department of Mathematical 
Sciences, UAE University,Al-Ain, P.O.B.17551, UAE,}

\affiliation{$^ {\bf \it b}$ Department of Mathematics,
  Ain Shams University, 11566,
 Cairo, Egypt}

\affiliation{$^ {\bf \it c}$ Physics Department,
United Arab Emirates University, Al-Ain, UAE. }

\date{\today}

\begin{abstract}
We study a simple two Higgs doublet model which reflects,
in a phenomenological way, the idea of compositeness for
the Higgs sector. It is relatively predictive. In one scenario, it
allows for a ``hidden" usual Higgs particle in the 100 GeV region
and a possible dark matter candidate.

\end{abstract}


\maketitle

\section{Introduction}

    Many fascinating models have been suggested
 for the Higgs sector
which is expected to be explored at
 the new CERN collider,
LHC in the next few years. 
These models are variously motivated by the ideas of 
supersymmetry \cite{susyreview}, 
possible  Nambu-Goldstone like (or "little")
 Higgs particles \cite{littlehreview},
possible extra dimensions \cite{exdimreview}  
, ``technicolor" binding composite
Higgs particles \cite{hs03} etc.

    Here we will be concerned with a model which may
  be related to the technicolor category
 but we will motivate
it from the assumption that the present Higgs model is not too 
far from being correct. Namely we want to abstract some properties
of the existing Higgs model and apply them to the two doublet
case.

    It is well known that the ordinary Higgs potential is 
formally
identical to the Gell-Mann Levy SU(2) linear sigma model
\cite{gl60} 
 potential:
\begin{equation}
   V =\alpha_1 I_1 + \alpha_3(I_1)^2,
\label{glpot}
\end{equation}
where the
 SU(2)$_L$ x SU(2)$_R$
 invariant $I_1$ is simply
expressed in terms of the scalar singlet $\sigma$
 and the pseudoscalar triplet
$ {\mbox{\boldmath ${\pi}$}} $ as
 $I_1 = \sigma^2 + {\mbox{\boldmath ${\pi}$}}^2$.
Of course the sigma is identified with the Higgs and the
$ {\mbox{\boldmath ${\pi}$}} $ with
 the particles eaten by the W and Z 
bosons.
The analogs of these two particles are the
 lowest lying ones
in ordinary QCD.
 Clearly a technicolor model
which is a straightforward copy of ordinary QCD would be
expected to give such a potential as a first approximation.
However, it is not easy to rigorously explore
the low lying spectrum of an arbitrary strongly interacting 
gauge theory \cite{sigt}.
Furthermore it is now known that a so-called
 ``walking" technicolor model \cite{fs08}
may be a more reasonable candidate than straightforwardly
extended QCD.
 Thus we will not insist that a technicolor induced
Higgs potential be identical to the above and shall
not try to estimate the particle masses. Rather we 
will just 
ask the effective Higgs
 potential to satisfy the general properties:

               1.  SU(2)$_L$ x SU(2)$_R$ flavor invariance.

               2. Parity invariance and charge conjugation
                  invariance.

   These are clearly very reasonable for a strong interaction
gauge theory with two massless flavors.
 
    In the present note we introduce a second Higgs doublet
based on the fact that the fundamental representation of SU(2)
is equivalent to its complex conjugate. This has the consequence
that the
 $( {\mbox{\boldmath ${\pi}$}} ,\sigma)$
 multiplet used above 
is
 irreducible under 
the chiral 
 SU(2)$_L$ x SU(2)$_R$ group
 without including the parity reversed
partners, denoted as $({\bf a},\eta)$. It seems natural to
investigate what happens when 
these parity reversed partners are 
included in a second Higgs doublet.
 Then, the three
basic invariants are, 
\begin{eqnarray}
 I_1 = \sigma^2 + {\mbox{\boldmath ${\pi}$}}^2,
\nonumber\\
 I_2 = \eta^2 + {\bf a}^2,
\nonumber\\
I_3=\sigma\eta- {\mbox{\boldmath ${\pi}$}}\cdot{\bf a}.
\label{invariants}
\end{eqnarray}
These forms are readily understandable since the 
two quartet fields may be regarded as 4-vectors in
the O(4)$\sim$ SU(2)$_L$xSU(2)$_R$ space \cite{sst81} and 
these
are the three basic invariants which can be made from them.
 Then the Higgs potential becomes,
\begin{equation}
V=\alpha_1 I_1+\alpha_2 I_2 +\alpha_3 I_1^2+
\alpha_4 I_2^2 +\alpha_5
I_3^2 +\alpha_6 I_1 I_2
\label{potential}
\end{equation}
The lack of terms linear in $I_3$ is due
 to the assumption of parity invariance.
This implies
that the fields {\bf a} and $\eta$ each only occur
in the potential paired off with either
itself or the other.
This feature may be expressed
 as the invariance of the 
potential
under the transformation:
\begin{equation}
\eta \rightarrow -\eta, \hspace{1cm}{\bf a}
\rightarrow -{\bf a},
\label{z2}
\end{equation} 
while the fields in the multiplet,
$( {\mbox{\boldmath ${\pi}$}} ,\sigma)$
are unchanged. 
Altogether there are six real constants.
 The present
 potential is supposed to be 
an effective one, arising
from some underlying renormalizable gauge theory.

     Interactions violating the invariances in 1. and 2.
 above are introduced as
perturbations in the model when the chiral fields are
coupled to the SU(2)xU(1) gauge fields in the usual way.
The two quartets of the chiral group are
 conveniently written for this purpose as two spinors,

\begin{eqnarray}
\Phi=
\left[
\begin{array}{c}
i\pi^{+}\\
\frac{\sigma-i\pi^0}{\sqrt{2}}
\end{array}
\right],
\hspace{0.5in}
\Psi=
\left[
\begin{array}{c}
-ia^{+}\\
\frac{\eta+i a^0}{\sqrt{2}}
\end{array}
\right],
\label{twodoublets}
\end{eqnarray}
and their conjugates.
Furthermore,
\begin{eqnarray}
\pi^+=\frac{\pi_1-i\pi_2}{\sqrt{2}}
 \hspace{0.5in}
a^+=\frac{a_1-ia_2}{\sqrt{2}} \label{api}.
\end{eqnarray}
     The gauged kinetic terms for these fields 
give the usual Lagrangian contribution:
\begin{equation}
{\cal L}=-D_{\mu}\Phi^{\dagger}D_{\mu} \Phi -
D_{\mu} \Psi^{\dagger} D_{\mu}  \Psi
\label{lagrangian}
\end{equation}
where
\begin{eqnarray}
&&D_{\mu}\Phi=\partial_{\mu}\Phi -ig W_{\mu}
 \Phi
 +\frac{ig'}{2} B_{\mu} 
\Phi,
\nonumber\\
&&D_{\mu}\Phi^{\dagger}=\partial_{\mu}
\Phi +ig  \Phi ^{\dagger}W_{\mu}
 -\frac{ig'}{2} B_{\mu} \Phi^{\dagger},
\label{covderiv}
\end{eqnarray}
with similar forms containing $\Psi$.
 Here $B_\mu$ is the U(1) gauge boson and the
SU(2) gauge bosons are expanded as:
\begin{eqnarray}
W_{\mu}=\frac{1}{2}{\mbox{\boldmath ${\tau}$}}^a
\cdot{\bf W}_{\mu}^a=
 \frac{1}{2}
 \left[
\begin{array}{cc}
W_{\mu}^{0}& \sqrt{2}W_{\mu}^+ \nonumber\\
\sqrt{2}W_{\mu}^-&-W_{\mu}^0.
\end{array}
\right] \label{gauge}
 \end{eqnarray}
The presence of the pure SU(2)xU(1) gauge field kinetic
terms in ${\cal L}$ is to be understood. Finally consider
the Yukawa terms containing the coupling of the quarks and 
leptons to the Higgs field. For this purpose, it seems
natural to demand the symmetry in Eq.(\ref{z2}), which can be 
rewritten as,
\begin{equation}
\Phi \rightarrow \Phi, \hspace{1cm} 
 \Psi \rightarrow -\Psi.
\label{spinorz2}
\end{equation}
 We also assume here that the quarks
 and leptons do not change under this symmetry
transformation. Then 
only the original Higgs multiplet $\Phi$ can couple to the 
fermions and the Yukawa couplings are just the usual ones.

\section{Discussion}

   Of course, there has been a very extensive discussion
of various two Higgs doublet models in the literature.
Recent related work includes that of Randall \cite{r07},
who considers a model with a heavy extra doublet
 in which the mixing 
 between singlet states is very small
(ie, large tan$\beta$), Ma \cite{m08} who
 stresses the connection with 
the dark matter problem,
 Gerard and Herquet \cite{gh07} who consider
connections with the custodial symmetry
and Lopez Honorez, Nezri, Oliver and Tytgat
 \cite{lh08}
 who discuss the dark matter
application extensively.

In the present work we emphasize that the idea
of compositeness for the Higgs bosons motivates both
the SU(2)$_L$ x SU(2)$_R$ 
as well as the P and C invariance of the Higgs potential.
This contains the usual custodial SU(2)$_V$ symmetry
together with a discrete $Z_2$ symmetry.
 If it is true that the model arises from some
underlying technicolor theory, it is reasonable to
think that the Higgs potential is an approximation
to the underlying theory
 describing the interactions of its lowest lying 
scalar
states. From this point of view the electroweak
interactions represent a perturbation to this ``strong"
interaction. Then it seems natural to classify the
symmetries of the Higgs potential according to the larger
``strong" interaction symmetry. This stands in contrast to
discussing the symmetry from the point of view of the spinors
$\Phi$ and $\Psi$ in Eq.(\ref{twodoublets}). In that language,
our invariant $I_1$ is identified as $2\Phi^{\dagger}\Phi$
while $I_2$ is identified as $2\Psi^{\dagger}\Psi$.
Also our  $I_3$ corresponds to the combination
 $[\Phi^{\dagger}\Psi 
 +\Psi^{\dagger}\Phi]$.
On the other hand, the combination
$i[\Phi^{\dagger}\Psi
 -\Psi^{\dagger}\Phi]$ is easily seen
to violate the proposed SU(2)$_L$ x SU(2)$_R$
invariance and will not be included. This
gives an additional simplification of the 
potential.

    It is interesting to remark that the
``minimal walking technicolor theory" \cite{ffrs}
automatically respects the symmetries 1. and 2.
which we are advocating. That theory contains
the Higgs bosons we are studying but also contains
other effective fields associated with the technicolor
interactions.

   In section III we will discuss 
 the potential part of this model and list all
 its terms. We will explicitly give the
expressions for the
scalar boson masses
 and two relevant coupling constants.
 In section IV we will explicitly
 give all the Lagrangian terms
for the interactions of the scalar bosons with the 
gauge bosons. The formulas in these sections can
be read to get an idea of the  interactions
of the extra bosons $\eta,a^0,a^{\pm}$ beyond the
 usual Higgs ($\sigma$).

    In the remainder of the paper, we discuss the possible
application of this model to ``shielding"
a relatively light ordinary Higgs boson from detection. This
is motivated from the unusually low value of its mass
expected from precision analyses of electroweak corrections.
Section V deals with a first model in which the
 decay of the usual Higgs into $\eta\eta$ is a competing mode
which might prevent seeing the Higgs in an
experiment which searches for
$b{\bar b}$ pairs in combination with a $\mu^+\mu^-$
Z boson indicator. The eta has no decay
 interactions in this model. That means that it
 could not ``hide" the Higgs in an experiment which just
 identifies events in which a Z is made 
 and no
other identifiable particles emerge. In section VI,
 we propose an alternate
way in the present framework to shield
 the Higgs from such an experiment 
too. 

\section{Higgs potential terms}
Recent discussions of general two Higgs
doublet potentials are given in \cite{pot}.
The present case, in which the field variables
comprise two O(4) vectors, is simpler than the general
case.
First, we note the constraints which follow
from the requirement that 
the Higgs potential be positive for large
 field configurations. This implies that the
quartic terms of the potential,
\begin{equation}
V=\cdots + \alpha_3(I_1)^2 + \alpha_4(I_2)^2
+[\alpha_5 \cos^2\theta+\alpha_6]I_1I_2,
\label{quarticV}
\end{equation}
where we used the O(4) property that
$I_3^2=I_1I_2\cos^2\theta$ for some angle
$\theta$, be positive for large field
 configurations.
Then taking either $I_1$ or $I_2$ to
 be dominant for large fields we get the
requirements:
\begin{equation}
  \alpha_3 > 0, \hspace{1cm} \alpha_4 > 0.
\label{alpha34}
\end{equation}
There is a possibility that $\alpha_5$
and/or  $\alpha_6$ 
may be negative. In such cases there is
an additional discriminant condition which is
obtained by
forbidding real roots of the quadratic form
obtained by dividing through by 
$(I_1)^2$. It has the form:
\begin{equation}
(\alpha_5cos^2\theta+\alpha_6)^2 <
4\alpha_3\alpha_4,
\label{disc}
\end{equation}
for any $\theta$. As examples,
\begin{equation}
(\alpha_5+\alpha_6)^2 <
4\alpha_3\alpha_4, \hspace{1cm}
\alpha_6^2 <
4\alpha_3\alpha_4.
\label{discexamples}
\end{equation}

Stronger information on the $\alpha$
coefficients arises, as to be discussed next,
 from calculating 
the particle
 masses
and interactions by expanding the potential
around the physical minimum $<\sigma> \ne 0$,
$<\eta> = 0$. The latter corresponds to our assumed
underlying parity invariance.
 A simple  
calculation verifies
 that
$<\partial V/\partial \sigma>$ =
$<\partial V/\partial \eta>$ = 0 for this minimum.

    The $\alpha_1$ and $\alpha_3$ terms
 in Eq.(\ref{potential}) correspond to the usual
single Higgs model. In the present case, 
 parity invariance prevents the $\sigma$ from mixing with
the $\eta$ so $\alpha_1$ and $\alpha_3$
are determined just as in the standard model.
Then $\alpha_1$ is negative and related to $\alpha_3$
by the minimization equation:
\begin{equation}
 \alpha_1 + 2\alpha_3v^2=0,
\label{min}
\end{equation}
where the vacuum value, v is given as
\begin{equation}
v=<\sigma>\approx 246 GeV.
\label{v}
\end{equation}

    The Higgs squared mass is obtained as
\begin{equation}
m_\sigma^2=8\alpha_3v^2.
\label{msigma}
\end{equation}
The potential also yields $m_{\mbox{\boldmath ${\pi}$}}^2 
=0$ for all three ``pions", which, in the unitary gauge
 get absorbed into massive gauge bosons.
For the particles in the $\Psi$ multiplet, the squared masses
are obtained as,
\begin{eqnarray}
m^2_\eta= 2\left[\alpha_2+(\alpha_5+\alpha_6)v^2\right],
\nonumber \\
m^2(a^0)=m^2(a^\pm)
\equiv m_a^2= 2\left[\alpha_2+\alpha_6v^2\right].
\label{aetamasses}
\end{eqnarray}
Notice that the three ``a" particles are degenerate in
mass. Furthermore there is no mixing between the two Higgs
multiplets.
    
Defining a shifted Higgs field $\sigma=v+{\tilde \sigma}$,
the interaction terms in the Lagrangian resulting from
the Higgs potential are:
\begin{eqnarray}
&&-\alpha_3({\tilde \sigma}^4 +4v{\tilde \sigma}^3)
-\alpha_4({\bf a}^2+\eta^2)^2
\nonumber \\
 &&-\alpha_5\eta^2(2v{\tilde \sigma}+ {\tilde \sigma}^2)
-\alpha_6({\bf a}^2+\eta^2)(2v{\tilde \sigma}+ {\tilde \sigma}^2).
\label{higgsinteractions}
\end{eqnarray}
 The interaction vertices for Feynman rules can be read off
from this equation.
For later convenience we identify the coupling constants for
 the $\sigma\eta\eta$ and ${\sigma}a^0a^0$ vertices,
\begin{equation}
g_{\sigma\eta\eta}=4v(\alpha_5+\alpha_6),
\hspace{1cm}g_{{\sigma}a^0a^0}=4v\alpha_6.
\label{g}
\end{equation}

It may be noted from Eqs.(\ref{min}) and (\ref{msigma})
 that specifying the Higgs mass, $m_\sigma$ will fix
the coefficients $\alpha_1$ and $\alpha_3$. Furthermore
specifying $m_\eta$, $m_a$ and $g_{\sigma\eta\eta}$ will
fix $\alpha_2$, $\alpha_5$ and $\alpha_6$. 
Information about $\alpha_4$
is related to the a-$\eta$ scattering amplitude.
We will not need  $\alpha_4$
in the present paper.

   The allowed ranges of the alpha parameters
are constrained by the requirement that the
squared masses $m_{\sigma}^2$,$m_{\eta}^2$
and $m_a^2$ be positive definite. This agrees
with the requirement that $V(\sigma,\eta)$
have a minimum, rather than a maximum or
saddle point at the point ($\sigma,\eta$)=
($v,0$). Specifically, we have:
\begin{eqnarray}
&&A \equiv \frac{{\partial}^{2}V}{\partial{\sigma}^2}(
v,0)= 2\alpha_1 +12v^2\alpha_3 =m_{\sigma}^2,
\nonumber \\
&&B \equiv  \frac{{\partial}^{2}V}{\partial{\sigma}
\partial\eta}(v,0)=0,
\nonumber \\
&&C\equiv \frac{{\partial}^{2}V}{\partial{\eta}^2}(
v,0)=2\alpha_2 +2v^2(\alpha_5+\alpha_6)=m_{\eta}^2.
\label{test}
\end{eqnarray}
The condition for no saddle point,
$B^2-AC<0$ as well the condition for
a minimum rather than a maximum,
 $A+C>0$ are both clearly satisfied
for positive definite squared masses.

\section{Gauge-Higgs interactions}

No undetermined parameters are introduced here.
It is necessary
to first give conventions for the $W_\mu^0$ - $B_\mu$
mixing matrix:

\begin{equation}
\left[
\begin{array}{c}  Z_\mu \\
                 A_\mu
\end{array}
\right]
=
\left[
\begin{array}{c c}
  c & s
\nonumber               \\
 - s  &  c
\end{array}
\right]
\left[
\begin{array}{c}
                        W_\mu^0 \\
                        B_\mu
\end{array}
\right],
\label{mixing}
\end{equation}
where $s$ and $c$ are respectively the sine and cosine
of the mixing angle. They are connected to the proton charge,
$e$ and the coupling constants in Eq.(\ref{covderiv}) by 
$g=-e/s$ and $g'=-e/c$.

   Since there are many terms we present them in four 
parts. First the terms in the Lagrangian containing a single
gauge boson are: 
\begin{eqnarray}
&& -ieA_\mu(a^-\stackrel{\leftrightarrow}{\partial_\mu}a^+)
-eZ_\mu\left[-\frac{i(c^2-s^2)}{2sc}(a^-
\stackrel{\leftrightarrow}{\partial_\mu}a^+)+\frac{1}{2sc}
(a^0\stackrel{\leftrightarrow}{\partial_\mu}\eta)\right]
\nonumber \\
&& -\frac{e}{2s}\left[W_\mu^+(a^-
\stackrel{\leftrightarrow}{\partial_\mu}
(\eta+ia^0))+
W_\mu^-(a^+
\stackrel{\leftrightarrow}{\partial_\mu}
(\eta-ia^0))\right].
\label{1gauge}
\end{eqnarray}

The terms involving both $W^+$ and $W^-$ are:
\begin{equation}
-W_\mu^+W_\mu^-\left[\frac{m_W^2}{v^2}(2v{\tilde \sigma}
+{\tilde \sigma}^2)+\frac{e^2}{s^2}(a^+a^-+\frac{\eta^2}{2}
+\frac{a^0a^0}{2})\right].
\label{2gaugeWW}
\end{equation}

The terms with two gauge bosons but no $W$'s are:
\begin{eqnarray}
&& -e^2A_{\mu}A_{\mu} a^+a^-
\nonumber \\
&&- Z_\mu Z_\mu\left[\frac{m_Z^2}{2v^2}
(2v{\tilde \sigma}+{\tilde \sigma}^2)
+\frac{e^2}{8c^2s^2}(\eta^2+a^0a^0)
+\frac{e^2(c^2-s^2)^2}{4s^2c^2}a^+a^-\right]
\nonumber \\
&&-A_{\mu}Z_{\mu}\frac{e^2(s^2-c^2)}{sc}a^+a^- .
\label{2gaugeZandA}
\end{eqnarray}

Finally, the terms with two gauge bosons
 containing a single $W$ take the form: 
\begin{equation}
-\frac{e^2}{2}(\frac{Z_\mu}{c}+\frac{A_\mu}{s})\left[
i\eta(a^+W_\mu^--a^-W_\mu^+) +a^0(a^+W_\mu^-+a^-W_\mu^+)
\right].
\label{other2gauge}
\end{equation}

    Feynman rules for gauge particle-Higgs particle
 vertices in the unitary gauge can be read off from the
above expressions. There is one observation about the
additional Higgs particles which is immediate. Since there
is a $Za^+a^-$ vertex and the width of the $Z$
is already well accounted for,  
 the $Z$ should not decay into
$a^+$ + $a^-$; this suggests considering
the mass range:
\begin{equation}
m_a > \frac{m_Z}{2}.
\label{amassbound}
\end{equation}
On the other hand there is no $Z\eta\eta$ vertex so
the $\eta$ mass has no lower bound from an analogous
decay. There is a $Za^0\eta$ vertex so the bound,
\begin{equation}
m_a +m_\eta > {m_Z},
\label{aetamassbound}
\end{equation}
is suggested. However,
this does not prevent $\eta$ from being 
very light if $a$ is of the order or somewhat
heavier than the $Z$. 

\section{First model for a hidden Higgs scenario}

Stimulated by precision calculations in the standard model
giving the Higgs mass prediction\cite{precision}, 
\begin{equation}
m_\sigma =89^{+38}_{-28} GeV,
\label{higgsfit}
\end{equation}
a number of groups have revived\cite{revival} an older idea
\cite{idea} that the Higgs might be light and not
yet detected because of a competitive
 decay mode to some hard to observe new particles. 
It would seem that a decay mode in the present model,
$\sigma\rightarrow\eta\eta$ is a reasonable candidate
for such a competing channel. As we observe above,
the $\eta$ occurs only in quadratic form in the Higgs
potential and only together with a conceivably much
heavier ${\bf a}$ particle in the gauge-Higgs part
of the Lagrangian. Thus it could have escaped
detection.

\begin{figure}[htbp]
\centering
\rotatebox{270}
{\includegraphics[width=7cm,clip=true]{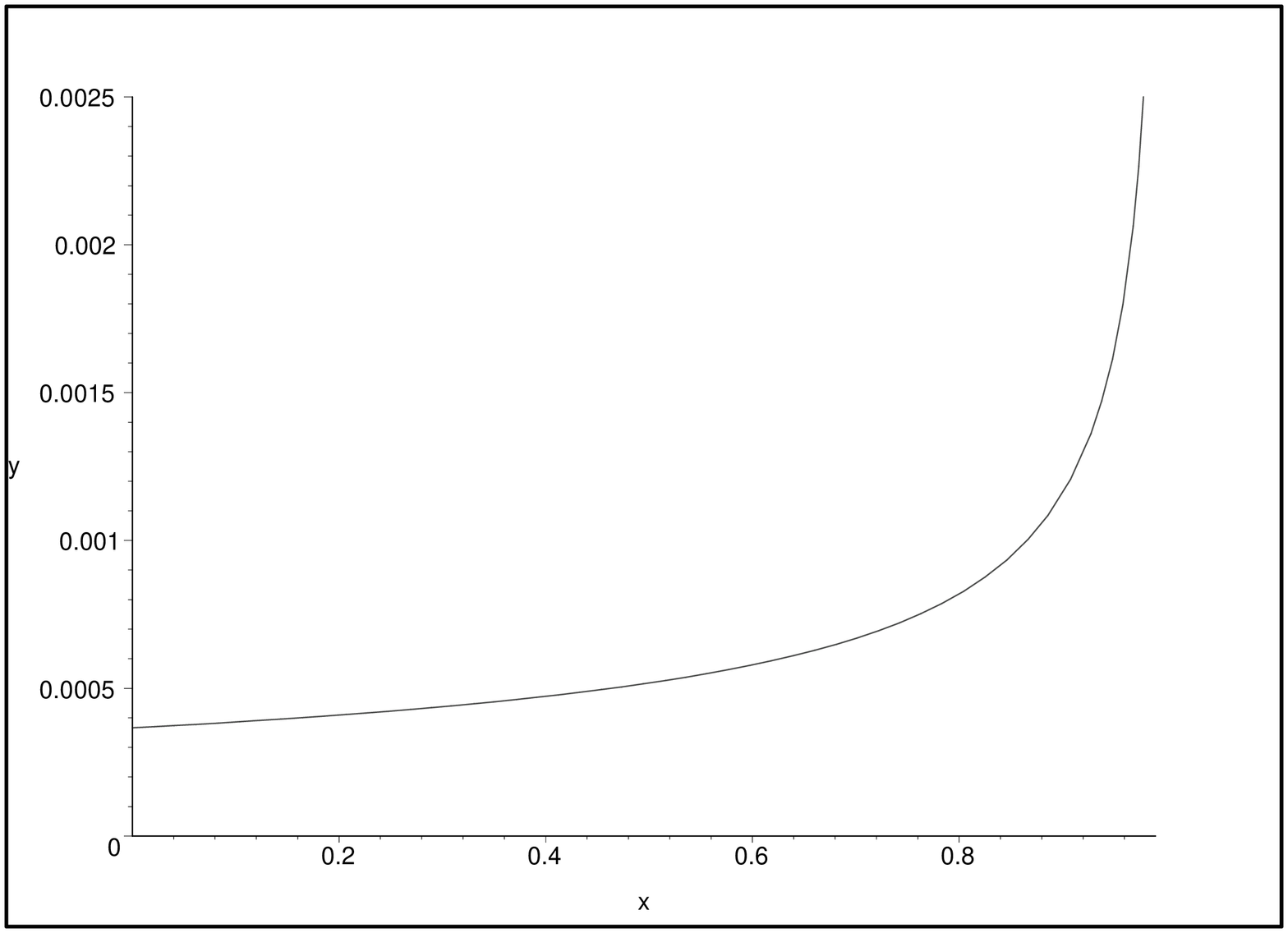}}
\caption[]{
$y$ vs.$x$.
}
\label{gvsmeta}
\end{figure}

     For the present purpose we need the formula for the
predicted Higgs width for its decay into $\eta\eta$:
\begin{equation}
\Gamma(\sigma\rightarrow \eta\eta)=
\frac{g_{\sigma\eta\eta}^2}{32\pi m_\sigma}
\sqrt{1-\frac{4m_\eta^2}{m_\sigma^2}},
\label{sigmatoetas}
\end{equation}
wherein $g_{\sigma\eta\eta}$ and $m_\eta$ are 
given in Eqs.(\ref{g}) and (\ref{aetamasses})respectively.
It can be seen that these two quantities are determined
by the parameters 
 $\alpha_2$ and $\alpha_5+\alpha_6$.
The typical Higgs search involves the reaction:
\begin{equation}
Z\rightarrow Z^* +\sigma,
\label{searchreaction}
\end{equation}
wherein the virtual $Z^*$ decays into $\mu^+\mu^-$
and the Higgs decays primarily into $b{\bar b}$ jets.
The formula for $\Gamma(\sigma\rightarrow b{\bar b})$ is:
\begin{equation}
\Gamma(\sigma\rightarrow b{\bar b})=
\frac{3m_\sigma m_b^2}{8\pi v^2}
\left(1-\frac{4m_b^2}{m_\sigma^2}\right)^{3/2},
\label{sigmabbar}
\end{equation}
where $m_b\approx$ 4.2 GeV is a conventional
estimate for the b quark mass. 
We need the ratio,
\begin{equation}
R=\frac{\Gamma(\sigma\rightarrow \eta\eta)}
{\Gamma(\sigma\rightarrow b{\bar b})}.
\label{ratio}
\end{equation}
Now if $P_{standard}$ gives the strength of the Higgs
signal in the standard model scenario, the reduced
strength due to the existence of the competitive
$\eta\eta$ decay mode in the present scenario would be,
\begin{eqnarray}
P_{new}&=&\frac{\Gamma(\sigma\rightarrow b{\bar 
b})}{\Gamma(\sigma\rightarrow b{\bar b})
+\Gamma(\sigma\rightarrow 
\eta\eta)}P_{standard}
\nonumber \\
&=&\frac{1}{1+R}P_{standard}.
\label{signaldecrease}\end{eqnarray}
It was noted \cite{revival} that a value,
$R=0.8$ would decrease the presently expected
Higgs signal below the detection threshold. 
Using the numbers just given we have,
\begin{equation}
R= 2184y\sqrt{1-x},
\label{R}
\end{equation}
where $x$=$(2m_\eta/m_\sigma)^2$
and $y$ = $(g_{\sigma\eta\eta}/v)^2$.
A plot of $y$ vs $x$ for the value 
R = 0.8 is shown in Fig.\ref{gvsmeta}.
Any point on that curve is a
solution for suppression of the $b{\bar b}$
Higgs signal.

 Three typical points, together
with the corresponding values of the 
Higgs potential parameters $\alpha_2$ and 
$\alpha_5 +\alpha_6$ are given in
 Table \ref{firsttable}.

\begin{table}[htbp]
\begin{center}
\begin{tabular}{c|c|c|c}
\hline 
  $m_{\eta}$ (GeV)  & $g_{\sigma\eta\eta}$ (GeV)&
 $\alpha_5+\alpha_6$ & $\alpha_2$ (GeV$^2$) \\
\hline \hline
14.1 & 4.8 & 4.91 x 10$^{-3}$   &-198      \\                  
31.5 &  5.6 & 5.69 x 10$^{-3}$  &+151       \\
42.2 &  8.4 &  8.51 x 10$^{-3}$ &+376       \\
\hline
\end{tabular}
\end{center}
\caption[]{Values of $m_\eta$, $g_{\sigma\eta\eta}$
and Higgs potential parameters which give suitable
suppression of the Higgs signal.
}    
 \label{firsttable}
\end{table}

    In the present scenario, with $m_a > m_\eta$, the $\eta$
boson has ``annihilation" modes but not decay modes. On the 
other hand the ${\bf a}$ particles have leading decay modes
 of the forms,
\begin{equation}
a^0\rightarrow Z + \eta, \hspace{1cm} 
a^+\rightarrow W^+ + \eta,
\label{adecays}
\end{equation}
wherein it has been assumed that the ${\bf a}$'s
are sufficiently heavier than the massive gauge bosons.
If the ${\bf a}$'s
are lighter than the massive gauge bosons
 but still heavier than the $\eta$, one would 
expect
important decays like,
\begin{equation}
a^0\rightarrow \eta + \mu^+ + \mu^-, \hspace{1cm}
a^+\rightarrow  \eta + \pi^+(139).
\label{lightadecays}
\end{equation}
These two decays are mediated by virtual $Z$
and $W$ bosons respectively.
The formula for the decay width of a heavy 
$ a^+$ by the reaction in Eq.(\ref{adecays})
 is  readily found to be:
\begin {equation}
\Gamma(a^+\rightarrow W^+ + \eta)=
\frac{k}{8{\pi}m_a^2}{\cal F}(a^+\rightarrow W^+ + \eta ),
\label{adecaywidth}
\end{equation}
where the momentum, {k} of each of the two
daughter particles  in the $a^+$
rest frame is:
\begin{equation}
k=\frac{1}{2m_a}\sqrt{[m_a^2-(m_\eta+m_W)^2]
[m_a^2-(m_\eta-m_W)^2]},
\label{k}
\end{equation}
and the squared amplitude summed over the final
$W^+$ polarization states is:
\begin{equation}
{\cal F}(a^+\rightarrow W^+ + \eta )=
(\frac{e}{2s})^2
[\frac{(m_\eta^2-m_a^2)^2}{m_W^2}-m_a^2-m_\eta^2
-2m_a\sqrt{k^2+m_\eta^2}].
\label{squareamp}
\end{equation}
We may use the same formula for 
$\Gamma(a^0\rightarrow Z + \eta)$ if we replace
$m_W$ by $m_Z$ and the overall factor $(e/(2s))^2$
by $(e/(2sc))^2$.
These ${\bf a}$ widths are listed in Table 
\ref{secondtable} for
a characteristic range of  ${\bf a}$ masses
in  cases where they are heavy enough
to decay into the gauge boson modes. The
$\eta$ mass is taken to be 31.5 GeV, the central
value in Table \ref{firsttable}.
It is seen that the widths are in the range
0.2 to 2 MeV for the ${\bf a}$ masses shown.
This may be compared to the width, 2.5 MeV,
 for the
Higgs (sigma) to decay into two $\eta$'s
according to Eq.(\ref{sigmatoetas}) taking
 $m_\eta$ =31.5 GeV.

 Also listed in Table \ref{secondtable} are the
  associated dimensionless coupling
constants $\alpha_5$ and $\alpha_6$
in the Higgs potential.
These are all less than unity, indicating
that for the mass range under discussion,
the new part of the Higgs sector is
not very ``strongly coupled".

\begin{table}[htbp]
\begin{center}
\begin{tabular}{c|c|c|c|c}
\hline 
  $m_{{\bf a}}$ (GeV)  & $\Gamma(a^+\rightarrow
W^+\eta$) (GeV)&
$\Gamma(a^0\rightarrow
Z\eta$) (GeV)&
 $\alpha_5$  & $\alpha_6$  \\
\hline \hline
150 &2.14 x 10$^{-4}$&1.52 x 10$^{-4}$    &-0.178  &0.235     \\                  
200 & 8.70 x 10$^{-4}$ &7.69 x 10$^{-4}$   &-0.322   & 0.379     \\
250  &  2.07 x 10$^{-3}$ &1.94 x 10$^{-3}$  &-0.508   &0.565      \\
\hline
\end{tabular}
\end{center}
\caption[]{
Widths of the ${\bf a}$ bosons for various mass values and
associated Higgs potential parameters.
}    
 \label{secondtable}
\end{table}

   It is amusing to remark that the quartic
 coupling constant $\alpha_5$ is negative. The
discussion at the end of section III implies that 
this is of no concern, since the squared masses
of all the Higgs particles are positive.
Note that the positive $\alpha_6$ is
 larger than the magnitude 
of $\alpha_5$. 

    Since the $\eta$ under study in the present
scenario does not have any decay modes, it
would appear to be another candidate for the
 ``dark matter"
required to understand galactic structures. Work
in this direction will be presented elsewhere.
 
\section{Second hidden Higgs model}

It was stressed in \cite{revival} that 
 Higgs search
experiments \cite{expts} which look for an appropriate Z 
(say by 
tagging $\mu^+\mu^-$pairs) together 
with the  $absence$
of any other particle signals
 could eliminate the possibility
of a light Higgs.
They point out that the Higgs can therefore
be shielded only if there is a ``cascade" decay
 of the decay
products ($\eta$'s in the first model) to  final
states containing a recognizable particle. The
$\eta$'s have no decays in our model, however.

\begin{figure}[htbp]
\centering
\rotatebox{270}
{\includegraphics[width=7cm,clip=true]{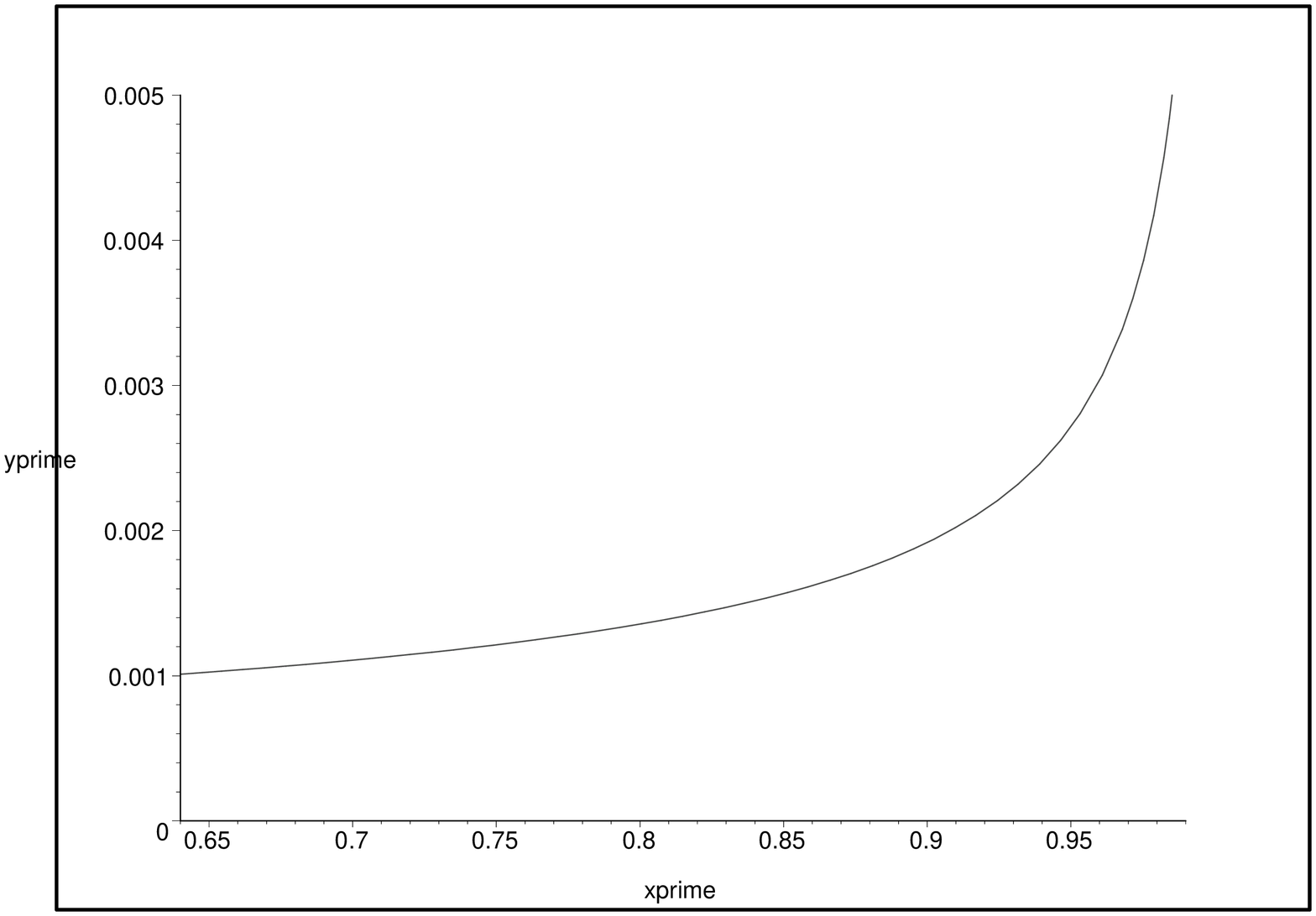}}
\caption[]{
$y'$ vs.$x'$.
}
\label{xprime}
\end{figure}

    We can shield a light Higgs in such an
experiment if we assume that
the three {\bf a} particles are lighter than half the Higgs 
mass and that the $\eta$ is lighter still. For
example, with a Higgs mass of 115 GeV,
 ${\bf a}$ masses of
 50 GeV could do the job. The  ${\bf a}$'s would be
heavy enough that they would not alter the
 well known Z width. (This mechanism is clearly suitable
for shielding Higgs bosons which
 are roughly more massive than the Z). Then the decay
modes
\begin{equation}
\sigma \rightarrow a^+ + a^-, \hspace{1cm}
\sigma \rightarrow a^0 + a^0,
\label{sigtoaa}
\end{equation}
are possible. 
Furthermore, the Eqs.(\ref{lightadecays})show that
the a's decay into the inert $\eta$ as well as
the recognizable particles
 $\pi^{\pm}$ or $\mu^+\mu^-$. It is still
 possible of course
for there to be some
 $\sigma \rightarrow \eta +\eta$ in addition
to these modes. To illustrate the present scenario
we will assume for simplicity that the 
coupling constant, $g_{\sigma\eta\eta}$ has
been tuned to be negligible.
 Then the relevant decay width is:
\begin{eqnarray}
&&\Gamma(\sigma\rightarrow a^+a^-)+
\Gamma(\sigma\rightarrow a^0a^0)=
\nonumber \\
&&3\Gamma(\sigma\rightarrow a^0a^0)=
\frac{3g_{\sigma{a^0}{a^0}}^2}{32\pi m_\sigma}
\sqrt{1-\frac{4m_a^2}{m_\sigma^2}}.
\label{sigmatoallas}
\end{eqnarray}
Proceeding as before we define,
\begin{equation}
R'=\frac{3\Gamma(\sigma\rightarrow a^0a^0)}
{\Gamma(\sigma\rightarrow b{\bar b})}=
1319y'\sqrt{1-x'^2},
\label{Rprime}
\end{equation}
where $x'$=$(2m_a/m_\sigma)^2$
and $y'$ = 3$(g_{\sigma{a^0a^0}}/v)^2$.
A plot of $y'$ vs $x'$ for the value 
$R'$ = 0.8 is shown in Fig.\ref{xprime}.
Any point on that curve is a
solution for suppression of the $b{\bar b}$
Higgs signal. In contrast to Fig.\ref{gvsmeta},
the $x'$ variable is not displayed down to zero, 
indicating that the shielding is only operative for
roughly $m_a>m_Z/2$.

 Three typical points, together
with the corresponding values of the 
Higgs potential parameters $\alpha_2$ and 
$\alpha_6$ are given in
 Table \ref{thirdtable}.

\begin{table}[htbp]
\begin{center}
\begin{tabular}{c|c|c|c}
\hline 
  $m_a$ (GeV)  & $g_{{\sigma}a^0a^0}$ (GeV)&
 $\alpha_6$ & $\alpha_2$ (GeV$^2$) \\
\hline \hline
48.1 & 4.7 & 4.82 x 10$^{-3}$   &576      \\                  
51.4 &  5.2 & 5.32 x 10$^{-3}$  &674       \\
54.5 &  6.2 &  6.32 x 10$^{-3}$ &723       \\
\hline
\end{tabular}
\end{center}
\caption[]{Values of $m_a$, $g_{{\sigma}a^0a^0}$
and Higgs potential parameters which give suitable
suppression of the Higgs signal. Here we take
$m_\sigma$=115 GeV.
}    
 \label{thirdtable}
\end{table}

One notices, as in the previous shielding model, that
the dimensionless coupling constant
 $\alpha_6$ is much less than 
one, so the Higgs bosons are not strongly coupled. 
If we want to tune the $\sigma\rightarrow\eta\eta$
 contribution
to be small, Eq.(\ref{g}) indicates
 that $\alpha_5$ should be taken
negative and slightly less in magnitude
 than $\alpha_6$. 

    Effectively, the present ``cascade" type shielding
 mechanism
would have characteristic signals of a
 $\pi^+\pi^-$ pair together with 
two unobservable $\eta$'s or two $\mu^+\mu^-$ pairs
together with two unobservable $\eta$'s.   

\section{Summary}
We noted that a technicolor theory underlying the
standard electroweak model is likely to result 
in a Higgs potential which posseses standard ``strong"
interaction symmetries like chiral SU(2),parity and
charge conjugation. This is obvious for the single Higgs
 doublet model. Imposing the same  requirement for a two
doublet model results in an interesting picture,
 which is rather constrained compared to a general two
doublet model.

In particular the second doublet doesn't mix with
 the first one although it interacts with it. This leads
to at least one possible dark matter candidate.

A number of very interesting Higgs scenarios can be
 constructed. The most conservative one would make the
second doublet heavier than the first. In
 this paper we considered an opposite picture with lighter
second doublet members. This provides extra
 decay modes for the usual 
Higgs boson and enables us to construct models which 
might hide the usual Higgs from being observed in 
certain experiments. These models involve all, but
one, of the parameters in our Higgs potential.
Information about the remaining one, $\alpha_4$ might
be found by considering the connection with dark matter
observations.
 
\section*{Acknowledgments} \vskip -.5cm We are
grateful to F. Sannino for reading
 the manuscript and pointing out that the
``minimal walking technicolor" model respects
the symmetries we are advocating here. We also
would like to thank
 A. Abdel-Rehim, D. Black, A.H. Fariborz and M.
Harada for helpful related discussions.
The work
of R.J. and J.S. is supported in part by the U.
S. DOE under Contract no. DE-FG-02-85ER 40231.


\begin{thebibliography}{10}
\bibitem{expts}A detailed discussion of the relevant
experiments is given in \cite{revival} above.
\end{thebibliography}
\end{document}